\documentclass[%
 reprint,
 amsmath,amssymb,
 aps,
pra,
]{revtex4-1}

\usepackage{graphicx}
\usepackage{dcolumn}
\usepackage{bm}
\usepackage{hyperref}
\hypersetup{colorlinks=true, citecolor=blue, urlcolor=blue, linkcolor=blue}
\usepackage{float}
\usepackage{color}
\usepackage[section]{placeins}
\usepackage{multirow}

\graphicspath{{figures/}}

\begin{document}

\preprint{APS/123-QED}

\title{Unbiased third-party bots lead to a tradeoff between cooperation and social payoffs}


\author{Zhixue He$^{1,2}$}
\author{Chen Shen$^3$}
\email{steven\_shen91@hotmail.com}
\author{Lei Shi$^1$}
\email{shi\_lei65@hotmail.com}
\author{Jun Tanimoto$^{2,3}$}

\affiliation{
\vspace{2mm}
\mbox{1. School of Statistics and Mathematics, Yunnan University of Finance and Economics, Kunming, 650221, China}
\mbox{2. Interdisciplinary Graduate School of Engineering Sciences, Kyushu University, Fukuoka, 816-8580, Japan}
\mbox{3. Faculty of Engineering Sciences, Kyushu University, Fukuoka,  816-8580, Japan}
}

\date{\today}

\begin{abstract}

The rise of artificial intelligence (AI) offers new opportunities to influence cooperative dynamics with greater applicability and control. In this paper, we examine the impact of third-party bots---agents that do not directly participate in games but unbiasedly modify the payoffs of normal players engaged in prisoner's dilemma interactions—--on the emergence of cooperation. Using an evolutionary simulation model, we demonstrate that unbiased bots are unable to shift the defective equilibrium among normal players in well-mixed populations. However, in structured populations, despite their unbiased actions, the bots spontaneously generate distinct impacts on cooperators and defectors, leading to enhanced cooperation. Notably, bots that apply negative influences are more effective at promoting cooperation than those applying positive ones, as fewer bots are needed to catalyze cooperative behavior among normal players. However, as the number of bots increases, a trade-off emerges: while cooperation is maintained, overall social payoffs decline. These findings highlight the need for careful management of AI's role in social systems, as even well-intentioned bots can have unintended consequences on collective outcomes.

\end{abstract}

\maketitle


\section{\label{sec:1}Introduction}

Committed individuals who consistently uphold their beliefs, opinions, and behaviors can exert significant influence on social dynamics\cite{shirado2017locally,mobilia2003does,aiyappa2024emergence}. These individuals can drive consensus in opinion dynamics\cite{mobilia2003does}, promote vaccination\cite{fukuda2016effects}, accelerate the emergence of social conventions\cite{centola2018experimental}, and even overturn established norms. They can also coordinate behaviors in contexts like social segregation\cite{jensen2018giant}. In a similar manner, with advancements in AI, bots are increasingly integrated into daily life, functioning as committed agents due to their controllability and scalability. For example, on platforms like Twitter, bots have been used to influence public opinion\cite{stella2018bots, bail2018exposure} and even affect elections\cite{Ferrara2017}. While ethical concerns surround their use, AI introduces new possibilities for shaping social dynamics, transcending traditional human limitations\cite{santos2024prosocial,rahwan2019machine}.

As AI continues to integrate into society, one of the most pressing social challenges that both humans and AI face is fostering cooperation\cite{santos2024prosocial,varol2017online,ishowo2019behavioural}. Despite its collective benefits, cooperation is difficult to establish among self-interested individuals who prioritize personal gains\cite{axelrod1981evolution, nowak2006five,wang2015universal}. Committed cooperators these who consistently choose to cooperate—may not change the noncooperative equilibrium in basic social dilemmas, they can still promote cooperation, even in one-shot, anonymous games, which represent some of the most challenging conditions for cooperation to emerge\cite{cardillo2020critical,shen2023committed}. In such scenarios, traditional reciprocity mechanisms\cite{nowak2006five}—direct, indirect, and network reciprocity, as well as kin and group selection—are absent. Despite the lack of these mechanisms, the presence of a minority of committed cooperators increases the probability that normal players encounter cooperators\cite{masuda2012evolution,shen2023committed}. When individuals place less emphasis on material payoffs, this heightened encounter probability leads to the promoted cooperation. Recent studies in behavioral economics confirm this effect, attributing it to the prosocial preferences inherent in human nature\cite{mao2017resilient}. This concept of committed individuals has been extended to human-machine interactions in various scenarios, including collective coordination\cite{shirado2017locally}, fairness issues\cite{santos2019evolution}, and the conundrum of prosocial punishment\cite{shen2024prosocial}. These studies demonstrate that incorporating committed bots can address challenges that may be difficult for humans to resolve on their own.

Unlike previous studies that treated committed individuals or bots as independent decision-makers, we consider a scenario where bots act as third-party regulators, influencing the payoffs of normal players without directly engaging in their interactions\cite{burt1995kinds, fehr2004third, halevy2015selfish}. For example, on platforms like Twitter, bots might amplify prosocial content, increasing visibility and engagement (positive payoff), or downrank harmful content, limiting its reach (negative payoff)\cite{soares2022spread,smith2022impact, hwang2024adopting}. Though these bots do not interact directly with players, they significantly shape social dynamics by externally modifying payoffs.

In our model, we extend one-shot and anonymous prisoner’s dilemma games (PDG) by incorporating third-party bot regulators. In this setup, normal players simultaneously choose between cooperating for collective benefit or defecting for self-interest, while unbiased bots influence player payoffs without directly engaging with them. Due to the simultaneous nature of the decision-making process and the anonymity of the players, the bots cannot distinguish between cooperators and defectors and instead apply their influence uniformly. Normal players update their strategies through social learning,\cite{szabo2007evolutionary,adami2016evolutionary,szabo2005phase} imitating strategies that yield higher payoffs, while the bots, acting like committed agents, do not change their actions over time.

Through our evolutionary simulation model, we explored the effects of third-party bots on cooperation outcomes. We find that these bots—whether exerting positive or negative influences on player payoffs—spontaneously generate distinct impacts on cooperators and defectors, despite their unbiased nature, leading to enhanced cooperation in networked populations but not in well-mixed populations. The enhancement effect is particularly strong when bots exert negative influences, as fewer bots are required to trigger cooperative behavior. However, in both cases of positive and negative influence, a trade-off emerges: while cooperation is maintained as the number of bots increases, overall social payoffs decline. Considering the positive assortativity\cite{grafen1979hawk,nemeth2010paradox,wang2015universal}, a common feature behind reciprocity mechanisms that promote cooperation, and the degree of asymmetric influence of these bots, we find that this tradeoff may be universal. These findings highlight the potential for even well-intentioned AI interventions to produce unintended consequences, emphasizing the need for careful management of AI in social systems.

\section{\label{sec:2}Model and method}

Our model employs a one-shot PD game in which paired players interact only once, making simultaneous decisions between cooperation ($C$) and defection ($D$). Mutual cooperation yields a reward $R$ for players, while mutual defection results in a punishment $P$. If a cooperator faces a defector, the defector gains a temptation payoff $T$, and the cooperator receives a sucker's payoff $S$. The dilemma arises under the conditions $2R>T+S$ and $T > R > P > S $, indicating that while mutual cooperation maximizes collective benefit, defection maximizes individual gain. We simplify by setting $ R = 1$, $P = 0$, $S = -r $, and $ T = 1 + r $, where $r > 1 $ represents the dilemma strength \cite{wang2015universal}, quantifying the relative advantage of defection over cooperation. The bots serve as third-party actors that does not directly participate in the PD game interactions with normal players but influences the players' payoffs by an amount of $|\beta|$. We focus on scenarios where the influence exerted by third-party bots does not exceed the payoff that individuals receive from mutual cooperation (i.e., $|\beta| < R$), ensuring that the bots' influence is not so substantial as to make mutual cooperation between individuals insignificant. Therefore, we limit the range of $\beta $ to (-1, 1).

We adopt two traditional population settings: structured and well-mixed. Considering a human-bot hybrid population of size $N$, with bots constituting a proportion $\rho$ and normal players $1 - \rho $. In the structured setting, both bots and players are randomly distributed on a grid lattice network with periodic boundaries, where bots and players interact with their 8-nearest neighbors (i.e., Moore neighborhood)\cite{szabo2007evolutionary}. Well-mixed population, equivalent to a fully connected network, players and bots interacts with $N-1$ others.

We obtain the results through agent-based simulations for structured population. Bots and players are initially randomly distributed on the network, with players having an equal probability of choosing either cooperation or defection. At each asynchronous time step, $N(1 - \rho)$ updates are performed, where in each update a player is selected randomly for strategy interaction and updates, ensuring that each player’s strategy is updated once on average. Strategy dynamics follow a social imitation rule with pairwise comparison, where social learning drives the preferential adoption of successful strategies through strategy dissemination \cite{szabo2007evolutionary,adami2016evolutionary,szabo2005phase}. The focal individual, says $i$ mimics the strategy of a randomly selected neighbor $j$ based on Fermi probability$W_{s_i \leftarrow  s_j}(\pi_i, \pi_i)$:
\begin{equation}
    W_{s_i \leftarrow  s_j}(\pi_i,\ \pi_j) = \frac{1}{ 1 + \exp\{ (\pi_i - \pi_j ) \kappa  \} }
\end{equation}
where $\pi_i$ is player $i$'s accumulative payoffs, which includes both the total payoffs derived from interactions with all neighboring players in the PD game and the influence of all neighboring bots on their payoffs. $\kappa$ represent the imitation strength. When $\kappa$ approaches 0, individuals engage in random imitation regardless of payoffs, i.e., weak imitation scenarios. Conversely, as $\kappa$ approaches $+\inf$, imitation is determined by payoff differences, representing strong imitation scenarios. Unless stated otherwise, $\kappa$ is set to 10 to investigate scenarios with strong imitation strength. Bots do not alter their behavior over time. For agent-based simulations within a structured population, we set $N=100^2$. To ensure stability, results are averaged over the last 5000 time steps from a total of $10^6$ steps. To mitigate the  interference caused by the random spatial distribution of individuals and bots, the final results are averaged over 50 independent simulations.

\begin{figure*}[!t]
 \centering
 \includegraphics[width=0.9\linewidth]{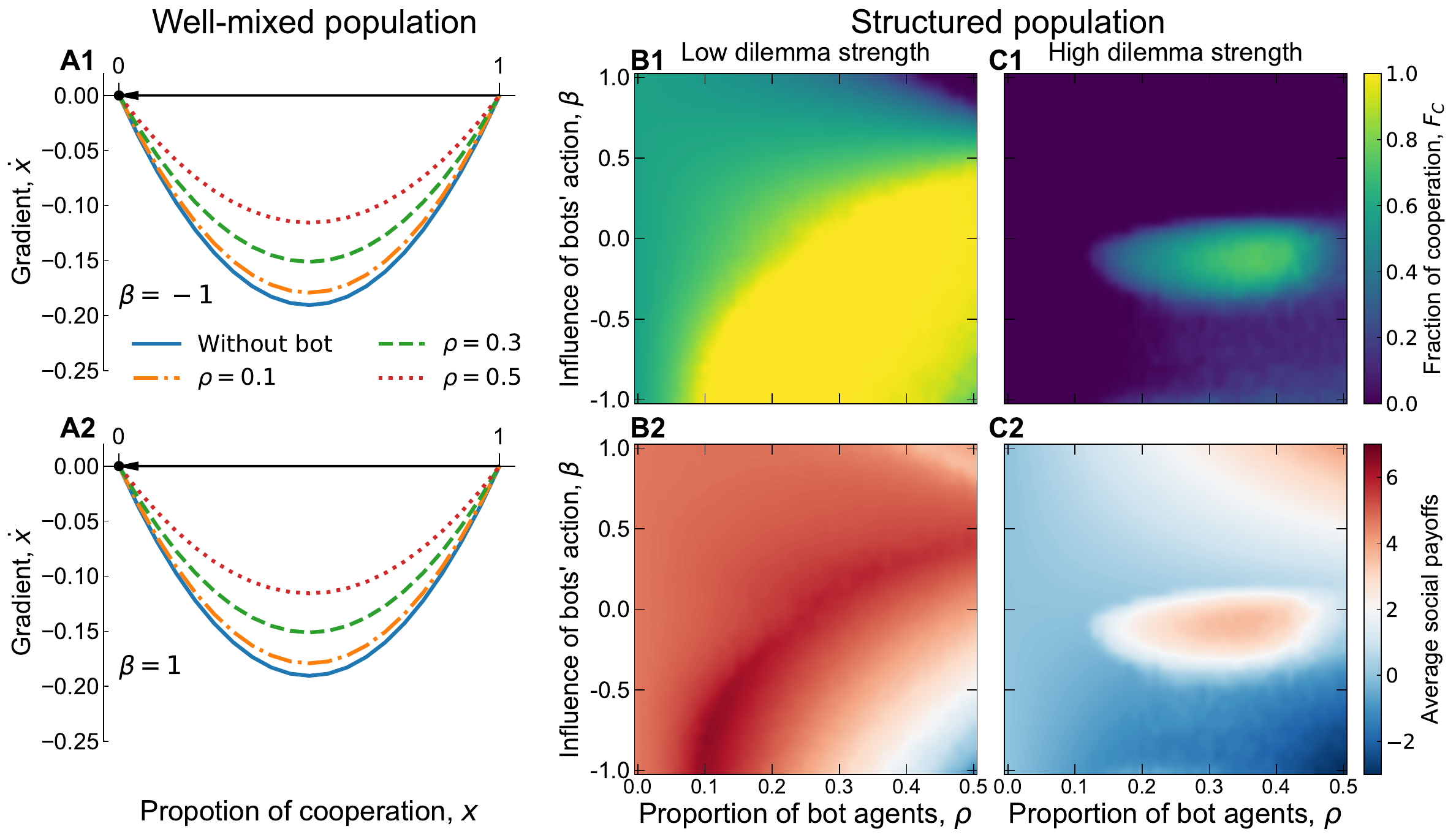}
 \caption{In structured populations, bots that impose negative effects are more effective at promoting cooperation than those impose positive effects. However, these bots introduce a trade-off: as their proportion increase, cooperation is further promoted, but at the cost of reduced social payoffs. Panel (A) illustrates the evolutionary dynamics of cooperation in a well-mixed population with $r = 0.2$, where neither positive nor negative influences from bots support the establishment of cooperation. The population converges to a state where $x=0$, i.e., pure defection. Results for the structured population depicted are the fraction of cooperation and average social payoffs among normal individuals as functions of the proportion of bots $\rho$ and influence of bot' action $\beta$ for (B) low-dilemma strength $r=0.05$ and (C) high-dilemma strength $r=0.14$ , respectively.
 }
 \label{fig:phas}
\end{figure*}

For an infinite well-mixed population, let $x$ denotes the proportion of cooperators among normal players. The payoffs for cooperators $\pi_C$ and defectors $\pi_D$ are calculated as follows:
\begin{equation}
    \begin{aligned}
    \pi_C &= \rho \beta + (1-\rho)[xR + (1-x)S] \\
    &= \rho \beta + (1-\rho)[x(1+r)-r] 
    \end{aligned},
\end{equation} 
\begin{equation}
    \begin{aligned}
    \pi_D &= \rho  \beta + (1-\rho)[xT + (1-x)P] \\
    &= \rho \beta + (1-\rho)(1+r)x
    \end{aligned}.
\end{equation}
In the social imitation dynamics, when a defector mimics a cooperator, it increases the number of cooperators by one, with a probability of $x^+$:
\begin{equation}
    x^+ = x(1-x)W_{D \leftarrow  C}(\pi_D,\ \pi_C) .
\end{equation}
 Conversely, when a cooperator mimics a defector, it decreases the number of cooperators by one, with a probability of $x^-$:
\begin{equation}
    x^- = x(1-x)W_{C \leftarrow  D}(\pi_C,\ \pi_D) . 
\end{equation}
The evolutionary dynamic of cooperation in the population \cite{nowak2006evolutionary}, denoted as $\dot{x}:=\frac{dx}{dt}$, is expressed as: 
\begin{equation}
\label{eqx}
    \begin{aligned}
    \dot{x} &= x^+ - x^-  \\
    & = x(1-x) ( W_{D \leftarrow  C}(\pi_D, \pi_C) - W_{C \leftarrow  D}(\pi_C, \pi_D))   
    \end{aligned}
\end{equation}
The equilibrium state of the system can obtained by numerically iterating this equations, initializing $ x \in (0, 1) $.

\section{\label{sec:3}Results}

Introducing bots into well-mixed populations affects the gradient of cooperation evolution but does not facilitate cooperation, regardless of whether the bots have a positive or negative impact on individuals, the system's equilibrium remains at pure defection(i.e., $x=0$ ), as illustrated in Fig.\ref{fig:phas}(A1) and (A2). Unlike well-mixed populations, structured populations exhibit network reciprocity\cite{nowak2006five,xia2023reputation}, which allows cooperators to form clusters, enabling them to resist the invasion of defectors. Yet, at a high dilemma strength, network reciprocity is insufficient to sustain cooperation \cite{wang2013insight}. This work investigates two representative scenarios: (i) network reciprocity successfully maintains cooperation at a low dilemma strength (i.e., $r=0.05$), and (ii) network reciprocity fails to avert the collapse of cooperation at a high dilemma strength (i.e., $r=0.14$).

\begin{figure*}[!t]
 \centering
 \includegraphics[width=0.95\linewidth]{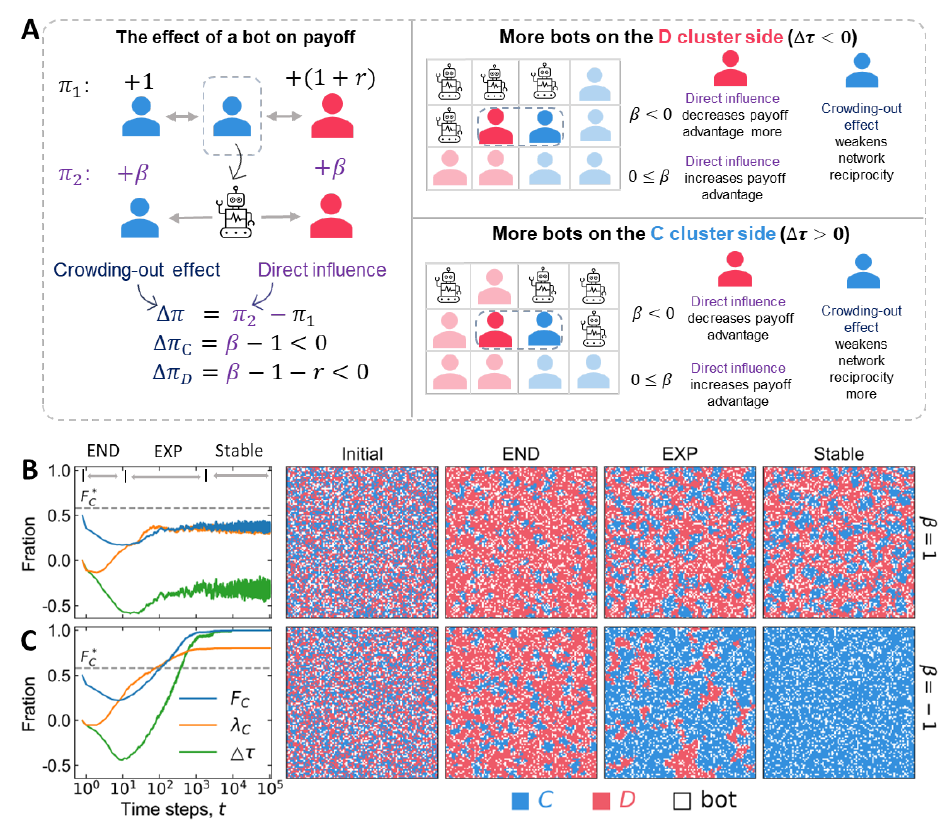}
 \caption{Depiced are 
 (A) the involvement of bots introduces both direct influences and crowding-out effect on individuals. The crowding-out effect arises when bots occupy the limited interaction space within a structured population, leading to a reduction in interactions between individuals and thereby diminishing individuals' potential maximum payoff by a $\Delta \pi $. Both cooperators and defectors are influenced by these effects, contingent upon the spatial arrangement of bots relative to the groups of cooperators and defectors. Both cooperators and defectors are impacted by these effects, but potential for cooperation to expand is contingent upon the spatial arrangement of bots with respect to the clusters of cooperators and defectors. (B) and (C) illustrate evolutionary snapshots under the influence of bots exerting positive (i.e., $\beta=1$) and negative impacts (i.e., $\beta=-1$), respectively, at $ r = 0.05 $. The evolution of cooperation in structured populations undergoes  enduring period (END) of discrete cooperator extinction, cooperative clusters expanding period (EXP), and eventual stabilization\cite{wang2013insight}. We reveal the self-organizing characteristics during the evolutionary process by analyzing the temporal dynamics of the cooperator fraction ($F_C\in[0,1]$), cooperator assortativity ($\lambda_C \in[-1,1]$), calculated as the mean difference between the proportion of cooperative and defective neighbors for cooperator, and bot-cooperator assortativity ($ \Delta \tau \in[-1,1]$), calculated as the mean difference in the proportion of cooperators and defectors surrounding the bots. The dashed line in left-most panels shows the steady-state fraction of cooperation $F_C^* = 0.58$ for traditional case without bots.
}
 \label{fig:snap1}
\end{figure*}

The influence of third-party bots on cooperation in structured populations depends on their actions. Under low dilemma strength, bots exerting a strong positive influence (e.g., $0.5<\beta$) provide direct benefits to normal players but paradoxically inhibit cooperation, as shown in Fig. \ref{fig:phas}(B1), where an increased bot proportion correlates with decreased cooperation. In contrast, for $0 < \beta < 0.5$, a moderate proportion of bots effectively enhances cooperation. In particular, even bots with a negative influence can significantly promote cooperation, with just a minority (e.g., $\rho=0.1$) sufficient to maintain high levels of cooperation at $\beta=-1$. However, bots also introduce a crowding-out effect, reducing the maximum potential payoff of neighboring individuals by $\Delta \beta$ due to their occupation of limited interaction space, as illustrated in Fig.~\ref{fig:snap1}(A). An increasing proportion of bots occupies more interaction space among normal individuals, diminishing overall societal payoffs, despite high cooperation levels being sustained for $0<\beta<1$, as seen in Fig.\ref{fig:phas}(B2). For bots with $-1<\beta<0$, their negative impact further decreases social payoffs.

\begin{figure*}[!t]
 \centering
 \includegraphics[width=0.9\linewidth]{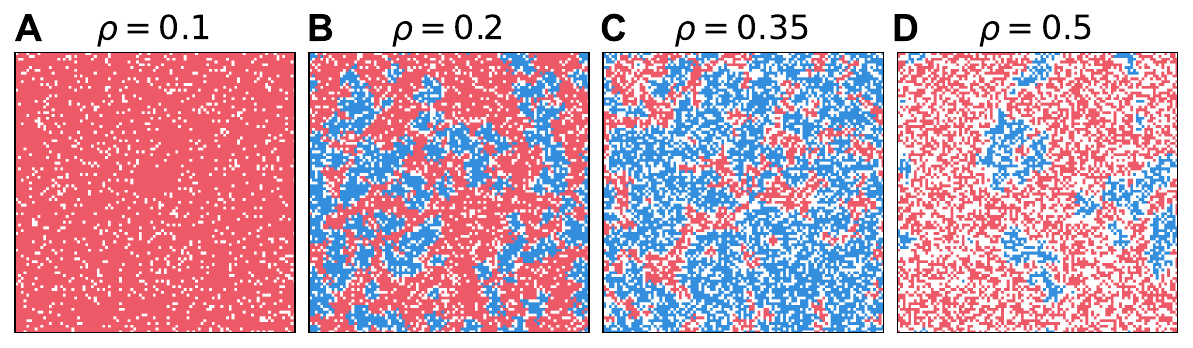}
 \caption{An intermediate proportion of bots optimizes cooperation facilitation, as too few bots are insufficient to establish cooperative clusters, while too many bots isolate individuals and hinder cluster expansion. The spatial distribution at steady state under high dilemma strength $r = 0.14 $ is shown for various bot proportions. Cooperators, defectors, and bots are depicted in blue, red, and white, respectively.}
 \label{fig:snap2}
\end{figure*}

\begin{figure*}[!t]
 \centering
 \includegraphics[width=0.75\linewidth]{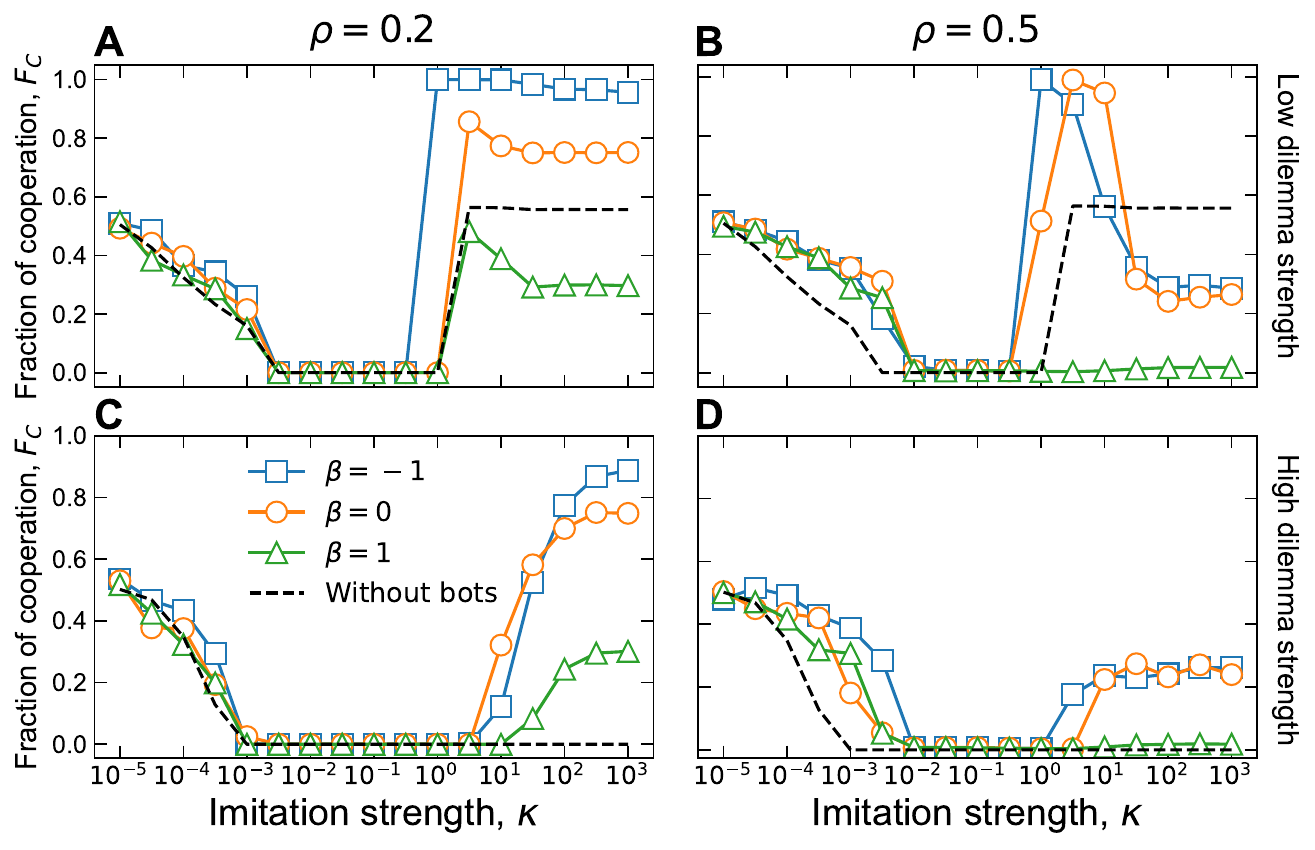}
 \caption{Under strong imitation conditions where individuals rely strictly on payoff differences, bots can cultivate and enhance cooperation. Depicted are the fraction of cooperation as a function of imitation strength $\kappa$ for low dilemma strength $r=0.05$ and high dilemma strength $r=0.14$.
 }
 \label{fig:k}
\end{figure*}

In contrast to low dilemma strength scenarios, where the optimal bot proportion can be determined across various $\beta$ values, high dilemma strength scenarios require a moderate density $\rho$ to sustain cooperation and achieve high payoffs, limited to a narrow range of $\beta$ values. While bots with a negative influence can help maintain cooperation, only those with a mild positive influence (i.e., $-0.28<\beta<0$) effectively sustain high cooperation levels, as shown in Fig. \ref{fig:phas}(C1). For $0.05<\beta<1$, bots fail to support cooperation, and in this parameter range, the absence of cooperation renders the crowding-out effect irrelevant; instead, their direct influence increases overall payoffs, as demonstrated in Fig. \ref{fig:phas}(C2). Similar to low dilemma strength scenarios, when bots exert a non-positive influence, increasing their proportion leads to a reduction in social payoffs. This highlights the trade-off between their ability to promote cooperation and the potential cost of reduced overall gains.

\begin{figure*}[!t]
 \centering
 \includegraphics[width=0.9\linewidth]{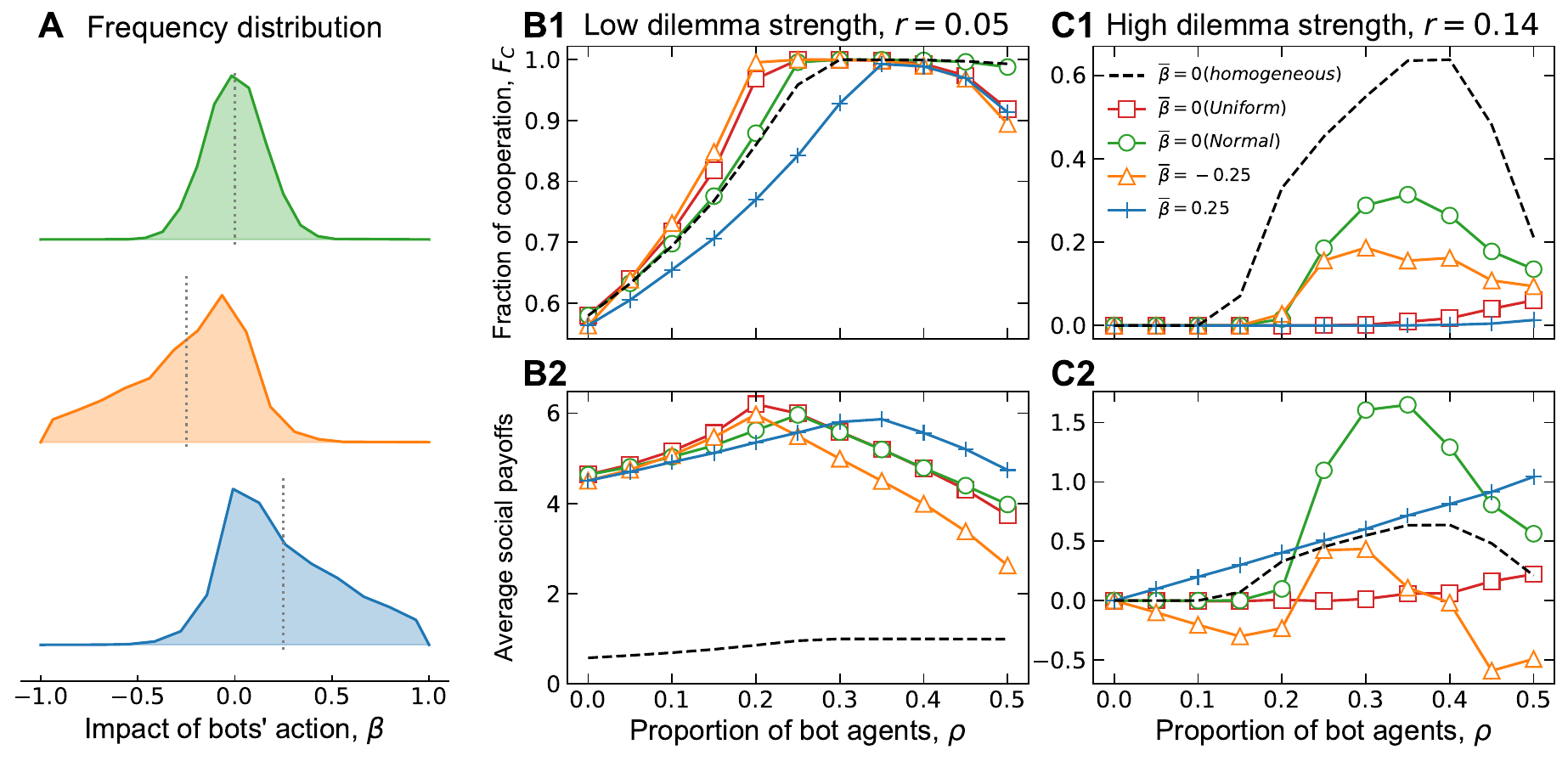}
 \caption{
 In heterogeneous action distribution scenarios, bots that exert non-positive influences continue to enhance cooperation, while those exerting positive influences not only fail to support cooperation but also undermine the efficacy of bots applying non-positive influences. The results consider bot behaviors with mean values $\overline{\beta}=0$ follows uniform and normal distribution, and two skewed distributions ($\overline{\beta}=-0.25$) and ($\overline{\beta}=0.25$). (A) shows the frequency distributions of bot behaviors under normal and skewed distributions. Behaviors following a normal distribution are sampled from $\mathcal{N}(0,\ 0.15)$, while skewed distributions are sampled from beta distribution $\mathcal{B}(1.2,\ 9.6)$ and rescaled to $[-1,1]$. (B) and (C) illustrate the fraction of cooperation $(F_C)$ and average social payoffs as functions of the proportion of bots $(\rho)$ for low-dilemma strength $r=0.05$ and high-dilemma strength $r=0.14$, respectively.
 }
 \label{fig:dis}
\end{figure*}

\begin{figure*}[!t]
 \centering
 \includegraphics[width=0.9\linewidth]{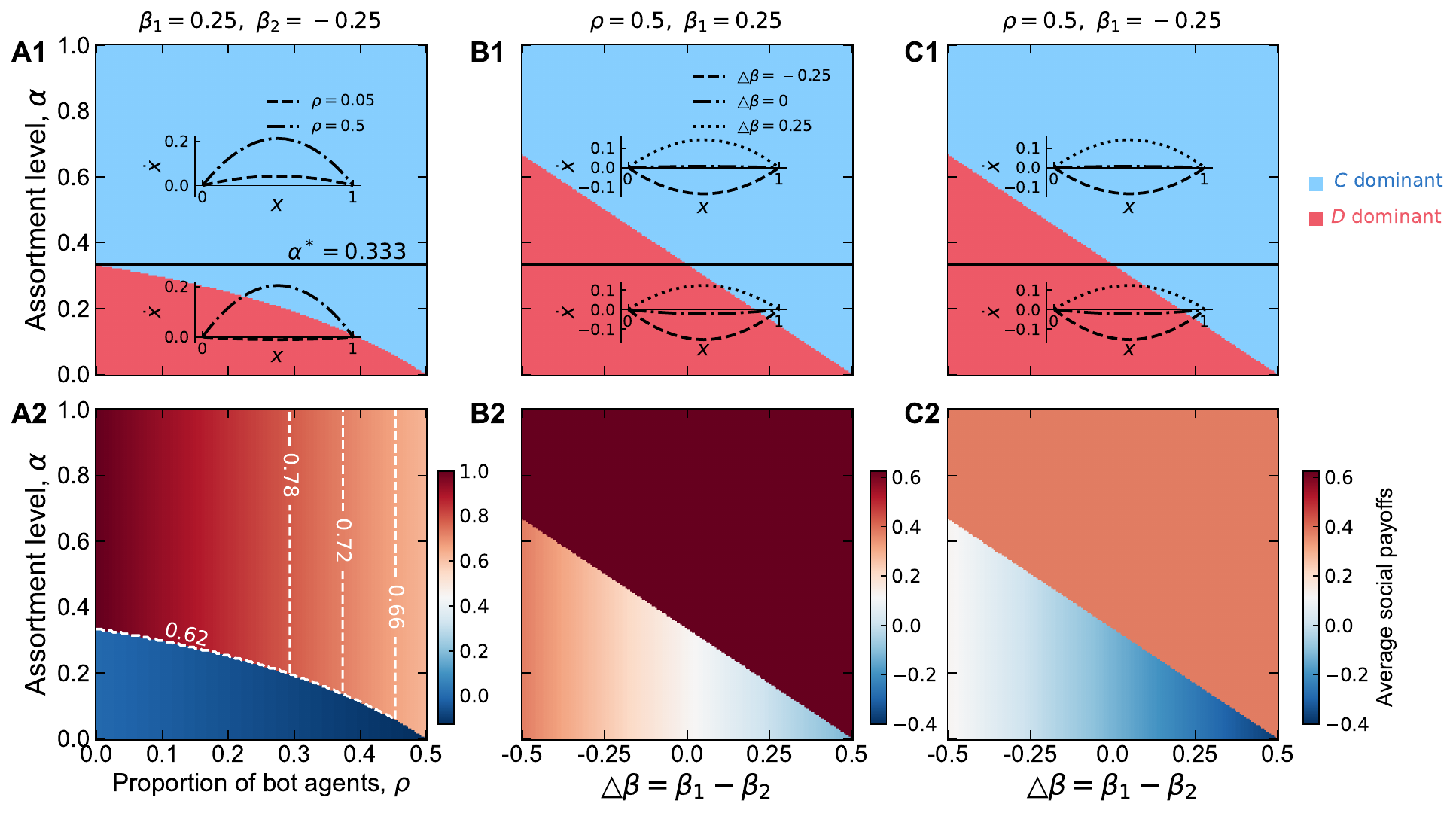}
 \caption{ In the presence of positive assortment, bots promote cooperation not through their positive or negative effects, but through their asymmetric influence on cooperators and defectors. While bots can support the establishment of cooperation, an increase in their numbers can reduce overall social payoffs. Depicted are strategy dominance (A1) and average social payoff (A2) as a function of the assortment level $\alpha$ and proportion of bots $\rho$. Blue regions indicate cooperation strategy dominance, while defection strategy dominant in red ones in (A1). The black solid line denotes the minimum critical assortment level $\alpha^* = 0.333$ for maintain cooperation without bots. Depicted are strategy dominance and average social payoffs as a function of the $\alpha$ and action influence variance $\Delta\beta = \beta_1 - \beta_2$. Cooperation dynamics for $\alpha = 0.34$ and $\alpha = 0.31$ are shown in the upper and lower panels of each subplot, respectively. The parameter $r$ is set to 0.5.}
 \label{fig:assort}
\end{figure*}

To further investigate these dynamics, we present the evolutionary outcomes in Fig.~\ref{fig:snap1}(B) and (C). The self-organizing process is characterized by cooperator assortativity ($\lambda_C \in [-1,1]$), which measures the average difference in the proportions of cooperative versus defective neighbors for cooperators \cite{fu2010invasion}, and bot-cooperator assortativity ($\Delta \tau \in [-1,1]$), assessing the average difference in the proportions of cooperators and defectors surrounding the bots. A $\lambda_C$ value approaching 1 indicates the formation of tight cooperator clusters, while $\Delta \tau<0$ suggests bots influence defectors more, and $\Delta \tau>0$ indicates a greater influence on cooperators.

Cooperation evolves through three periods \cite{wang2013insight}: (i) the early enduring period (END), where most isolated cooperators vanish, leaving a few small clusters; (ii) the expanding period (EXP), where these clusters grow; and (iii) the stabilization period. During the END period, $C$ clusters experience reduced payoffs due to the crowding-out effect of surrounding bots, making them more susceptible to defector invasion and weakening network reciprocity. As most cooperators disappear, bots exert greater influence over defectors, resulting in $\Delta \tau<0$. In the EXP period, cooperator clusters expand, increasing $\lambda_C$. However, in the scenario with $\beta = 1$, most bots enhance defector payoffs rather than supporting cooperators, as indicated by the persistently negative $\Delta \tau<0$, which impedes the expansion of cooperator clusters. Thus, bots lead to a lower steady-state cooperation level compared to traditional scenarios without bots, as shown in Fig.~\ref{fig:snap1}(B). Conversely, when $\beta = -1$, bots reduce defector payoffs with minimal impact on cooperators. This is evident in the corresponding END and EXP snapshots in Fig.~\ref{fig:snap1}(C), where bots within defector clusters act like Trojan horses, facilitating the expansion of cooperation and ultimately achieving full cooperation. 

The proportion of bots is crucial for their effectiveness in promoting cooperation. An insufficient proportion fails to reduce defector payoffs and support cooperation clusters, as shown in the leftmost panel of Fig.~\ref{fig:snap2}. Conversely, too many bots occupy limited interaction space, isolating individuals and hindering the expansion of $C$ clusters, as illustrated in the rightmost panel of Fig.~\ref{fig:snap2}.

We examine how varying imitation strengths affect bots' influence on players with different levels of payoff dependency, as shown in Fig.~\ref{fig:k}. When $\kappa \rightarrow 0$, players imitate randomly, regardless of payoffs. In low imitation strength scenarios ($\kappa < 10^{-3}$), bots have a negligible impact on imitation dynamics. At moderate imitation strengths ($10^{-3} < \kappa < 1$), randomness in imitation hinders cooperation, and bots do not improve this situation. At high imitation strengths ($1 < \kappa$), while all types of bots are ineffective in promoting cooperation for $\rho = 0.5$ and $r = 0.05$, their influence remains consistent under other conditions. Unlike zealots, whose influence is limited under weak imitation strength\cite{cardillo2020critical}, third-party bots effectively maintain cooperation under strong imitation strength.

To understand the impact of bots' actions, we investigate scenarios with heterogeneous bot behaviors, allowing for both positive and negative influences within the population. We analyze bot actions using uniform distributions, normal distributions sampled from $\mathcal{N}(0, 0.15)$, and skewed distributions from a beta distribution $\mathcal{B}(1.2, 9.6)$, rescaled to $[-1, 1]$ to achieve means $\overline{\beta}$ of -0.25 and 0.25. Fig.~\ref{fig:dis}(A) shows the action frequency distributions of these bots. Under low dilemma strength, normal and uniform distributions yield better outcomes than a homogeneous distribution with a mean of 0 ($\overline{\beta} = 0$), as demonstrated in Fig.~\ref{fig:dis}(B1). However, under high dilemma strength, bots with positive impacts fail to sustain cooperation and weaken the effectiveness of bots with non-positive impacts. Fig.~\ref{fig:dis}(C1) illustrates that cooperation levels are diminished due to non-positive influences, which hinder cooperation maintenance. In a skewed distribution with $\overline{\beta} = -0.25$, the proportion of bots effectively promoting cooperation ($-0.3 < \beta \leq 0$) is lower than in the normal distribution scenario, resulting in a reduced $F_C$. For a skewed distribution with $\overline{\beta} = 0.25$, the large number of positively influencing bots cannot sustain cooperation, making the impact of non-positive bots negligible. Additionally, there exists an optimal moderate proportion of bots that can enhance cooperation while ensuring high payoff levels, as shown in Fig.~\ref{fig:dis}(B2) and (C2).

One-shot and anonymous interactions are rare in reality, but this artificial setting allows us to solely focus on the impact of third-party regulator bots on cooperation among normal players. To broaden the applicability of our findings, we extend our analysis to include other reciprocity mechanisms—such as direct reciprocity, indirect reciprocity, group selection, kin selection, and network reciprocity—by leveraging positive assortment, where cooperators are more likely to interact with other cooperators, a key feature across these mechanisms \cite{taylor2007transforming}. In scenarios involving network reciprocity, the bots, although unbiased, spontaneously create distinct effects on cooperators and defectors. This suggests that third-party regulator bots may similarly generate differential impacts on cooperators and defectors when integrated with other reciprocity mechanisms, possibly due to underlying factors that have not yet been explored. Thus, we generalize our results by allowing bots to exert distinct effects on cooperators and defectors, without specifying the precise mechanisms by which these distinctions arise. The influence of bots on cooperator is denoted by $\beta_1$ and on defectors by $\beta_2$, with the difference in impact given by $\Delta \beta = \beta_1 - \beta_2$. In models with positive assortment \cite{grafen1979hawk,nemeth2010paradox,wang2015universal}, the probability of players interacting with those using the same strategy is denoted by $\alpha$; otherwise, interactions are random. The average payoffs for cooperators and defectors are calculated as follows:
\begin{equation}
    \label{eqc}
    \begin{aligned}
    \pi_C &= \rho \beta_1 + (1-\rho)[  \alpha R + (1-\alpha) (xR + (1-x)S ) ] \\
    &= \rho \beta_1 + (1-\rho)[  \alpha + (1-\alpha) (x + xr -r ) ]
    \end{aligned},
\end{equation} 
\begin{equation}
    \label{eqd}
    \begin{aligned}
    \pi_D &= \rho \beta_2 + (1-\rho)[  \alpha P + (1-\alpha) (xT + (1-x)P ) ] \\
    &= \rho \beta_2 + (1-\rho)(1-\alpha)(1+r )x
    \end{aligned}.
\end{equation}
The evolutionary dynamic can be derived using equation (\ref{eqx}), substituting the relevant payoffs from equation (\ref{eqc}) and (\ref{eqd}).


The results in Fig.~\ref{fig:assort}(A1) demonstrate that while bots can enhance cooperation under positive assortment, but their also reduces overall social payoffs. Consistent with previous findings, there is an optimal proportion of bots that maximizes both cooperation and payoff levels. Interestingly, when bots are capable of targeted actions, the ability of bots to promote cooperation depends on the relative magnitude of their influence on cooperators versus defectors, rather than whether their influence is positive or negative, as shown in Fig.~\ref{fig:assort}(B1) and (C1). If bots can exert positive asymmetric effects (i.e., $\Delta \beta = \beta_1 - \beta_2 > 0$), such as providing greater benefits to cooperators than to defectors, or if their negative impact on cooperators is less severe than that on defectors, cooperation can be promoted; otherwise, they counterproductively inhibit cooperation and weaken the assortment. Moreover, bots exerting positive influences (i.e., $\beta_1=0.25$) sustain higher social payoffs compared to those exerting negative influences (i.e., $\beta_1=-0.25$), as illustrated in Fig.~\ref{fig:assort}(B2) and (C2). Importantly, the trade-off created by bots may be widespread. As shown in Fig.~\ref{fig:assort}(A1) and (A2), while an increase in the proportion of bots can establish cooperation, further increases leads to a decline in overall social payoffs.

\section{\label{sec:4}Discussions}
As artificial intelligence becomes more integrated into society, the scalability and controllability of bots offer new avenues for addressing social challenges that are difficult for humans alone, such as combating misinformation \cite{ciampaglia2018research}, curbing rumor spread \cite{jones2024containing}, and reducing opinion conflicts \cite{flamino2023political} in human-bot hybrid interactions \cite{shirado2017locally,rahwan2019machine,varol2017online,ishowo2019behavioural,shen2024prosocial,guo2023facilitating,birhane2023science}. Unlike previous approaches that focus on bots as anonymous agents interacting directly with humans \cite{shirado2017locally,shen2024prosocial,guo2023facilitating}, this study integrates third-party bots into a population of normal players engaged in a prisoner's dilemma game, where players gain material payoffs and update strategies through social learning. These bots do not participate as game players but consistently exert uniform influence on the payoffs of normal players, without distinguishing between cooperators and defectors, due to the one-shot, anonymous, simultaneous nature of the game.

Our findings reveal that in structured populations, third-party bots, although unbiased, can spontaneously exert different effects on cooperators and defectors, leading to enhanced cooperation. In contrast, this effect is absent in well-mixed populations, where the bots fail to produce significant differences in the behavior of normal players. Notably, bots applying negative influence—such as reducing payoffs—are more effective at promoting cooperation than those offering positive rewards; indeed, bots acting as "good guys" by providing large rewards may unintentionally suppress cooperation. Even inactive bots (i.e., $\beta=0$) can facilitate cooperation by functioning like empty nodes that isolate interactions and promote cooperative behavior \cite{alizon2008empty}. However, regardless of their specific influence, bots introduce a crowding-out effect that reduces overall social welfare as their numbers increase. This presents a trade-off: while bots can enhance cooperation, an excessive number of them can diminish social payoffs. These results underscore the importance of carefully managing the behavior and deployment of bots in digital governance to avoid unintended consequences, even with well-meaning interventions.

The ability of unbiased third-party bots to promote cooperation may also extend to scenarios involving other reciprocity mechanisms. In network reciprocity, bots impact cooperators and defectors differently by occupying network nodes, influencing the formation of cooperative clusters, and creating a crowding-out effect. This effect does not rely on specific network configurations, making it robust across different network structures. Similarly, this mechanism may apply to direct reciprocity, indirect reciprocity, group selection, and kin selection, as they all share the common feature of mutual recognition of cooperators, known as positive assortment. This inherent recognition may allow unbiased third-party bots to enhance cooperation in these scenarios as well. For instance, in direct reciprocity, bots might identify individuals through repeated interactions \cite{roberts2008evolution}; in indirect reciprocity, they could use social reputation mechanisms to differentiate actions based on reputational profiles \cite{schmid2021unified}. Based on this assumption, we show that third-party bots can enhance cooperation, but the trade-off between increased cooperation and reduced social welfare persists under positive assortment, suggesting this may be a widespread consequence of bot implementation. Our findings offer an initial insight into the effectiveness of bots within reciprocal frameworks. Since different reciprocity mechanisms influence cooperation through distinct pathways, further research is needed to explore how unbiased bots can spontaneously exert asymmetric effects on cooperators and defectors across these mechanisms.

Human decision-making is a critical aspect in exploring the influence of bots within human-bot interactions. In social dilemmas, self-interested individuals pose challenging barriers to the emergence of cooperation, as they prioritize personal gain over collective benefit \cite{szabo2007evolutionary, adami2016evolutionary}. Analyzing the effects of bots on such individuals can reveal the effectiveness of bots in addressing social dilemmas. However, human decisions are influenced not only by self-interest but also by social norms, preferences, and moral standards \cite{ostrom2000collective, capraro2021mathematical}. This raises a pertinent question for future research: how do bots impact human adherence to social norms and moral principles? Furthermore, individuals may hold biases against bots \cite{karpus2021algorithm, kaur2022trustworthy}, and negative behaviors exhibited by bots—similar to punishment—could trigger retaliatory actions due to negative emotions \cite{jensen2010punishment}. Understanding these psychological effects is essential for a comprehensive grasp of how bots shape collective behavior.

In conclusion, our findings demonstrate that the impact of committed minority extends beyond direct interactions. On the other hand, our research reveals how the actions and proportions of bots influence social cooperation and payoffs, which can provide insights for designing human-bot systems that leverage committed individuals to address social dilemmas.

\paragraph*{Acknowledgments}
We acknowledge the support provided by (i) the National Natural Science Foundation of China (Grant No.11931015), Major Program of National Fund of Philosophy and Social Science of China (Grants Nos.~22\&ZD158 and ~22VRCO49) to L.S.; (ii) China Scholarship Council (Grant No. ~202308530309) to Z.\,H; 
(iii) JSPS KAKENHI (Grant no. JP 23H03499) to C.\,S., and (iv) the grant-in-Aid for Scientific Research from JSPS, Japan, KAKENHI (Grant No. JP 20H02314 and JP 23H03499) awarded to J.\,T.

\paragraph*{Author contributions}
Z.H. and C.S. conceived, designed the study and performed research; Z.H., C.S. L.S. and J.T. analyzed results and wrote manuscript; L.S. and J.T. reviewed and provided supervision.

\paragraph*{Competing interest} Authors declare that they have no conflict of interests.

\paragraph*{Data availability}
The data and code that support the findings of this study are available from the corresponding authors upon reasonable request.



\bibliographystyle{unsrt}
\bibliography{ref}


\end{document}